\documentclass{article}
\usepackage{spconf,amsmath,graphicx}

\usepackage{subfigure} 
\graphicspath{{./image/}}

\usepackage{amssymb}

\DeclareMathOperator*{\argmax}{argmax}

\usepackage{epstopdf}
\usepackage{tabularx}
\usepackage{multirow}
\usepackage{textcomp}
\usepackage{url}
\usepackage{lipsum}
\usepackage{makecell}

\usepackage{bm}
\usepackage{calc}

\usepackage{color}
\usepackage{dsfont}

\usepackage{algorithm}
\usepackage[noend]{algpseudocode}

\title{Multi-view Audio and Music Classification}
\name{\begin{tabular}{c}Huy Phan$^{*1}$, Huy Le Nguyen$^{2}$, Oliver Y. Ch\'{e}n$^{3}$, Lam Pham$^{4}$, Philipp Koch$^{5}$, \\ Ian McLoughlin$^{6}$, Alfred Mertins$^{5}$ \end{tabular}}
\address{$^1$Queen Mary University of London, UK, {~~~~~}$^2$HCMC University of Technology, Vietnam \\
	$^3$University of Oxford, UK, {~~~~~} $^4$Austrian Institute of Technology Vienna, Austria  \\ 
	$^5$University of L\"ubeck, Germany, {~~~~~} $^6$Singapore Institute of Technology, Singapore \\
	{$^\ast$Correspondence email: \tt h.phan@qmul.ac.uk} 
}

\begin{document}
	 \ninept
	\maketitle
	\begin{abstract}
	We propose in this work a multi-view learning approach for audio and music classification. Considering four typical low-level representations (i.e. different views) commonly used for audio and music recognition tasks, the proposed multi-view network consists of four subnetworks, each handling one input types. The learned embedding in the subnetworks are then concatenated to form the multi-view embedding for classification similar to a simple concatenation network. However, apart from the joint classification branch, the network also maintains four classification branches on the single-view embedding of the subnetworks. A novel method is then proposed to keep track of the learning behavior on the classification branches and adapt their weights to proportionally blend their gradients for network training. The weights are adapted in such a way that learning on a branch that is generalizing well will be encouraged whereas learning on a branch that is overfitting will be slowed down. Experiments on three different audio and music classification tasks show that the proposed multi-view network not only outperforms the single-view baselines but also is superior to the multi-view baselines based on concatenation and late fusion.
	\end{abstract}
\vspace{-0.15cm}
\section{Introduction}
\vspace{-0.15cm}
Good embeddings are crucial for machine learning tasks \cite{Pascual2019, Phan2017, Cramer2019}.  For audio and music classification, in particular, such an embedding can be learned from a variety of low-level features which have been developed alongside the development of the research field, such as Mel-scaled spectrogram \cite{Phaye2019,Phan2019b, Piczak2015,Choi2017b}, Gammatone spectrogram \cite{Phan2019,Zhang2019,Phan2017}, Constant-Q transform (CQT) spectrogram \cite{Sigtia2016, Huzaifah2017, Lidy2016}, and even raw waveform \cite{Dai2017,Tokozume2018}. Oftentimes, recognition results obtained from embeddings learned from different low-level inputs vary in the sense that one embedding is good for some target classes while another is good for some other target classes. This implies that the embeddings are complement and the low-level inputs can be reasonably considered as different views of the target data. Intuitively, owing to their complementarity, jointly learning from these views should leverage their individual strength and gives rise to performance gain on a task at hand \cite{McLoughlin2020, Pham2020}. However, it is not always the case in practice as a naive fusion scheme, e.g. concatenation \cite{Phan2017, Wang2020, Phan2020}, may result in performance degradation rather than improvement, i.e. the multi-view performance could be worse than that of the best single view \cite{ Wang2020, Phan2020}. The reason is that different single-view subnetworks learn at different rates and converge/overfit at different times during the training course. As a result, fusing out-of-sync single-view subnetworks via concatenation results in a suboptimal multi-view model. Late fusion is another common approach for fusing information from multiple views; however, separate training single-view networks is unable to take into account interaction between the views.

Inspired by prior work in \cite{Wang2020, Phan2020}, we propose a novel multi-view learning method based on deep learning for audio and music classification that overcomes the aforementioned issues. In the proposed approach, a multi-view network is designed so that we are able to gain assess to the convergence/overfitting behavior of the constituent single-view subnetworks. This then allows us to individualize their learning during the training process. In intuitive, learning on subnetworks that are genaralizing well is encouraged whereas learning on subnetworks that are ovefitting is slowed down. This is accomplished by assigning different weights to the subnetworks' losses prior to blending their gradients \cite{Wang2020, Phan2020}. The weights are adaptively adjusted according to the subnetworks' learning behavior. By doing this, we are able to regulate the contribution of each view into the multi-view embedding rather than even their contribution as in the case of simple concatenation. %Employing four common low-level audio features (Mel-scale spectrogram, Gammatone spectrogram, CQT spectrogram, and raw waveform) as four single views of a target audio data, 
Our experiments on three different audio and music classification tasks (environmental sound classification, audio scene classification, and music genre classification) show that the multi-view embedding learned via the proposed method consistently results in better performance than that obtained by all the single-view baselines and the multi-view baselines based on concatenation and late fusion.

\vspace{-0.15cm}
\section{Learning multi-view audio/music embedding}
\vspace{-0.15cm}
\subsection{Network architecture}
\begin{figure} [!t]
	\centering
	\includegraphics[width=0.85\linewidth]{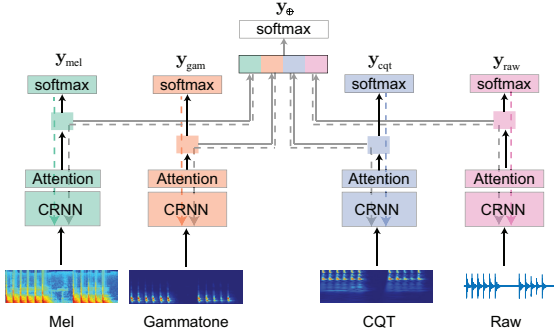}
	\vspace{-0.25cm}
	\caption{\small Illustration of the network for learning multi-view embedding. The dash lines represent the gradient backpropagation flows.}
	\label{fig:multiview_network}
	\vspace{-0.2cm}
\end{figure}

\vspace{-0.15cm}
We adopt four low-level features, including Mel-scale spectrogram, Gammatone spectrogram, CQT spectrogram, and raw waveform, which are most widely used for audio and music analysis under deep learning paradigms. They are considered as different views of the underlying data distribution of a audio/music classification task at hand.  The proposed network for learning multi-view embedding is illustrated in Fig.~\ref{fig:multiview_network}. It consists of four subnetworks, each of which is to process one of the low-level inputs. The multi-view embedding is formed by concatenating the embeddings learned by the view-specific subnetworks. However, apart from the concatenation branch, the view-specific CRNNs also maintain their own classification branches which serve as a gateway to access their learning behavior. The subnetworks are realized by convolutional recurrent neural networks (CRNNs) that are described below.

{\bf 2D CRNNs:} The 2D inputs (i.e. Mel-scale, Gammatone, and CQT spectrogram) have a general size of $T \times F$ where $T$ is the number of time frames and $F$ frequency bands. 
The CRNNs corresponding to the 2D inputs share a similar network architecture whose configuration is shown in Table \ref{tab:2dcrnn}. The architecture features six convolutional layers, 
each associated with Rectified Linear Unit (ReLU) activation \cite{Nair2010}, batch normalization \cite{Ioffe2015}, and a max pooling layer. The max pooling layers have a common kernel size of $2\times1$ and stride $1\times1$ to reduce size of spectral dimension by half while maintaining the temporal dimension of the input. On top of the convolutional layers, a bidirectional recurrent neural network (biRNN) layer is employed for sequential modelling on the time dimension. It is realized by Gated Recurrent Units (GRUs) \cite{Cho2014}.
  The sequence of recurrent outputs is then reduced to a feature vector (i.e. view-specific embedding) via spatio-temporal attention pooling suggested in \cite{Phan2019}. For classification purpose, the 2D CRNNs make use of two fully-connected layers with 
  ReLU activation, followed by a final output layer with softmax. 
  A dropout rate of $0.1$ is applied to the convolutional layers, the recurrent layers, and the fully-connected layers. 

{\bf 1D CRNN:} %The raw waveform input to the 1D CRNN has length of $66,650$ samples (roughly 1.5s at 44.1 kHz sampling rate). 
The 1D CRNN's configuration is shown in Table~\ref{tab:1dcrnn}. Out of eight convolutional layers, the first two (\emph{conv01} and \emph{conv02}) coupled with the max pooling layer (\emph{pool02}) are tailored to transform the raw input into a 2D representation as in \cite{Tokozume2018,Dai2017}. The rest of the network can then be parametrized similar to the 2D CRNNs described above, except for the \emph{conv1} and \emph{pool1} for which larger (temporal) kernel sizes are set to efficiently deal with the larger (temporal) input as well as to shorten its temporal dimension. 

Beside the single-view classification branches on the CRNNs, classification on the multi-view embedding is carried out via two fully-connected layers with 4096 hidden units and ReLU activation, followed by a final output layer with softmax. Let $\mathbf{y}$ denote the one-hot encoding ground-truth and $\mathbf{\hat{y}}^{(k)}$ denote the output of the classification branch $k \in \{mel, gam, cqt, raw, \oplus\}$. We use $\oplus$ to denote the multi-view classification branch.  The cross-entropy loss induced by the branch $k$ on a set of $M$ samples reads:
\begin{align}
\small
\mathcal{L}^{(k)} = -\frac{1}{M}\sum\nolimits_{m=1}^M\mathbf{y}_m\log(\mathbf{\hat{y}}_{m}^{(k)}).
\end{align}
The total loss used for training at the training step $n$ is computed by:
\begin{align}
\small
\mathcal{L}(n) = \sum\nolimits_kw^{(k)}(n)\mathcal{L}^{(k)}(n),
\end{align}
where $w^{(k)}(n)$ denotes the weight of the classification branch $k$ at the training time $n$. $w^{(k)}(n)$ is adapted during the training process according to the learning behavior of the branch $k$.

\setlength\tabcolsep{2.25pt}
\begin{table}[t!]
	\scriptsize
	\caption{\small Configuration of the 2D CRNNs. The output shape is of the format $(time, frequency, channel)$. Here, the number of kernels $F_l = (32, 64, 128, 128, 256, 512)$ for six convolutional layers indexed by $l = (1, 2 , 3, 4, 5, 6)$.}
	\vspace{-0.2cm}
	\begin{center}
		\begin{tabular}{>{\arraybackslash}m{0.4in}>{\centering\arraybackslash}m{0.4in}>{\centering\arraybackslash}m{0.35in}>{\centering\arraybackslash}m{0.4in}>{\centering\arraybackslash}m{0.35in}>{\centering\arraybackslash}m{1in}}
			\hline
			{Layer}  & {Filter size} & Stride & {\#filters} & {Padding} & {Output shape} \parbox{0pt}{\rule{0.pt}{1ex+\baselineskip}}\\ [0ex] 	% inserts table heading
			\hline
			Input &  & & & & $(T, 64, 1)$ \parbox{0.5pt}{\rule{0pt}{0ex+\baselineskip}}\\ [0ex] 	% inserts
			\hline
			conv-$l$ & $(3, 3)$ & $(1, 1)$ & $F_l$ & SAME & $(T , \frac{64}{2^{l-1}} , F_l)$ \parbox{0.5pt}{\rule{0pt}{0ex+\baselineskip}}\\ [0ex] 	% inserts
			pool-$l$  & $(1 , 2)$ & $(1, 1)$& &VALID & $(T , \frac{64}{2^{l}} , F_l)$ \parbox{0.5pt}{\rule{0pt}{0ex+\baselineskip}}\\ [0ex] 	% inserts
			\hline
			reshape & & && & $(T, 512)$ \parbox{0.5pt}{\rule{0pt}{0ex+\baselineskip}}\\ [0ex] 	% inserts
			\hline
			biRNN & && $2\cdot256$ & & $(T, 512)$ \parbox{0.5pt}{\rule{0pt}{0ex+\baselineskip}}\\ [0ex] 	% inserts
			\hline
			attention & && $64$ & & $(512,)$ \parbox{0.5pt}{\rule{0pt}{0ex+\baselineskip}}\\ [0ex] 	% inserts
			\hline
			fc1 & && $1024$ & & $(1024,)$ \parbox{0.5pt}{\rule{0pt}{0ex+\baselineskip}}\\ [0ex] 	% inserts
			fc2 && & $1024$ & & $(1024,)$ \parbox{0.5pt}{\rule{0pt}{0ex+\baselineskip}}\\ [0ex] 	% inserts
			fc3 & && \#classes & & $(\text{\#classes},)$ \parbox{0.5pt}{\rule{0pt}{0ex+\baselineskip}}\\ [0ex] 	% inserts
			\hline
		\end{tabular}
	\end{center}
	\label{tab:2dcrnn}
	\vspace{-0.6cm}
\end{table}

\setlength\tabcolsep{2.25pt}
\begin{table}[t!]
	\scriptsize
	\caption{\small Configuration of the 1D CRNN. The output shape is of the format $(time, frequency, channel)$. Here, the number of kernels $F_l = (64, 128, 128, 256, 512)$ for five convolutional layers indexed by $l = (2, 3, 4, 5, 6)$.}
	\vspace{-0.2cm}
	\begin{center}
		\begin{tabular}{>{\arraybackslash}m{0.4in}>{\centering\arraybackslash}m{0.4in}>{\centering\arraybackslash}m{0.35in}>{\centering\arraybackslash}m{0.4in}>{\centering\arraybackslash}m{0.35in}>{\centering\arraybackslash}m{1in}}
			\hline
			{Layer}  & {Filter size} & Stride & {\#filters} & {Padding} & {Output shape} \parbox{0pt}{\rule{0.pt}{1ex+\baselineskip}}\\ [0ex] 	% inserts table heading
			\hline
			Input &  & & & & $(66,650, 1, 1)$ \parbox{0.5pt}{\rule{0pt}{0ex+\baselineskip}}\\ [0ex] 	% inserts
			\hline
			conv01 & $(64, 1)$ & $(2, 1)$ & $32$ & VALID &  \parbox{0.5pt}{\rule{0pt}{0ex+\baselineskip}}\\ [0ex] 	% inserts
			conv02 & $(16, 1)$ & $(2, 1)$ & $64$ & VALID &  $(16,640, 1, 64)$\parbox{0.5pt}{\rule{0pt}{0ex+\baselineskip}}\\ [0ex] 	% inserts
			pool02  & $(64, 1)$ & $(64, 1)$&& VALID & $(260 , 1, 64)$ \parbox{0.5pt}{\rule{0pt}{0ex+\baselineskip}}\\ [0ex] 	% inserts
			\hline
			reshape & & && & $(260, 64, 1)$ \parbox{0.5pt}{\rule{0pt}{0ex+\baselineskip}}\\ [0ex] 	% inserts
			\hline
			conv1  & $(5, 3)$ &$(1, 1)$& $32$ & SAME &$(260 , 64 , 32)$ \parbox{0.5pt}{\rule{0pt}{0ex+\baselineskip}}\\ [0ex] 	% inserts
			pool1  & $(4, 2)$  &  $(4, 2)$&& VALID & $(65, 32, 32)$ \parbox{0.5pt}{\rule{0pt}{0ex+\baselineskip}}\\ [0ex] 	% inserts
			\hline
			conv-$l$ & $(3, 3)$ & $(1, 1)$ & $F_l$ & SAME & $(65 , \frac{64}{2^{l-1}} , F_l)$ \parbox{0.5pt}{\rule{0pt}{0ex+\baselineskip}}\\ [0ex] 	% inserts
			pool-$l$  & $(1 , 2)$ & $(1, 1)$& &VALID & $(65 , \frac{64}{2^{l}} , F_l)$ \parbox{0.5pt}{\rule{0pt}{0ex+\baselineskip}}\\ [0ex] 	% inserts
			%\hline
			\hline
			reshape & & && & $(65, 512)$ \parbox{0.5pt}{\rule{0pt}{0ex+\baselineskip}}\\ [0ex] 	% inserts
			\hline
			biRNN & && $2\cdot256$ & & $(65, 512)$ \parbox{0.5pt}{\rule{0pt}{0ex+\baselineskip}}\\ [0ex] 	% inserts
			\hline
			attention & && $64$ & & $(512,)$ \parbox{0.5pt}{\rule{0pt}{0ex+\baselineskip}}\\ [0ex] 	% inserts
			\hline
			fc1 & && $1024$ & & $(1024,)$ \parbox{0.5pt}{\rule{0pt}{0ex+\baselineskip}}\\ [0ex] 	% inserts
			fc2 && & $1024$ & & $(1024,)$ \parbox{0.5pt}{\rule{0pt}{0ex+\baselineskip}}\\ [0ex] 	% inserts
			fc3 & && \#classes & & $(\text{\#classes},)$ \parbox{0.5pt}{\rule{0pt}{0ex+\baselineskip}}\\ [0ex] 	% inserts
			\hline
		\end{tabular}
	\end{center}
	\label{tab:1dcrnn}
	\vspace{-0.6cm}
\end{table}

\vspace{-0.15cm}
\subsection{Adaptive gradient blending}
\vspace{-0.15cm}
Similar to \cite{Wang2020,Phan2020}, learning behavior on the branch $k$ can be assessed via the generalization measure $G^{(k)}$ and the overfitting measure $O^{(k)}$. In intuition, $G^{(k)}$ represents the information about the target distribution gained via training and $O^{(k)}$ represents the gap between information gain on the training set and the target distribution. $G^{(k)}$ and $O^{(k)}$ at the training step $n$ are approximated as:
\begin{align}
\small
G^{(k)}(n) &\approx L^{*(k)}_\diamond -  L^{(k)}_\diamond(n), \label{eq:G} \\
O^{(k)}(n) &\approx (L^{*(k)}_{tr} -  L^{(k)}_{tr}(n)) - (L^{*(k)}_\diamond -  L^{(k)}_\diamond(n)). \label{eq:O} 
\end{align}
In (\ref{eq:G}) and (\ref{eq:O}), $L^{(k)}_{tr}(n)$ and $L^{(k)}_\diamond(n)$ denote the loss on a training set and the loss on a test set (i.e. the true loss) at the training step $n$, respectively. $L^{*(k)}_{tr}$ and $L^{*(k)}_\diamond$ denote the training and true loss references, respectively. Since the true loss is unknown, we approximate it by the loss on a validation set. The weight $w^{(k)}$ for the branch $k$ is then computed as the ratio of generalization and overfitting measure:
\begin{align}
\small
w^{(k)}(n) = \frac{1}{Z} \frac{G^{(k)}(n)}{O^{(k)2}(n)}, \label{eq:w}
\end{align}
where $Z$ is a normalization factor. A network branch which is generalizing (i.e., large $G^{k}$ and small $O^{k}$) will have a larger weight to encourage its learning. In contrast, a network branch which is overfitting (i.e., small $G^{k}$ and large $O^{k}$) will have a smaller weight to discourage its learning. A square for $O^{k}$ in (\ref{eq:w}) is to avoid the situation when an underfitting network branch still scores very well on the generalization-over-overfitting ratio and receives a large weight.
\begin{algorithm}[t!]
	\small
	\caption{Computation of an adaptive weight}
	\begin{algorithmic}[1]
		
		\Procedure{AdaptiveWeight}{$L_{tr}, L_\diamond, L^*_{tr}, L^*_\diamond, W$}       %\Comment{This is a test}
		\State {\bf Input:} $L_{tr}[1\ldots n]$: list of training loss values \\
		{~~~~~~~~~~~~~~~~~~}$L_\diamond[1\ldots n]$: list of true loss values \\
		{~~~~~~~~~~~~~~~~~~}$L^*_{tr}$: current best training loss value\\
		{~~~~~~~~~~~~~~~~~~}$L^*_\diamond$: current best true loss value \\
		{~~~~~~~~~~~~~~~~~~}$W$: smoothing window size
		\State {\bf Output:} $w(n)$: weight at the training time $n$
		\State $\bar{L}_{tr}(n)\!=\!\!\text{mean}(\!L_{tr}[(n\!\!-\!\!W)\ldots n])$ \Comment{Smoothed training loss}
		\State $\bar{L}_\diamond(n)\!=\!\!\text{mean}(L_\diamond[(n\!\!-\!\!W)\ldots n])$\Comment{Smoothed true loss}
		\State $G(n) = L^*_\diamond - \bar{L}_\diamond(n)$ \Comment{Eq. (\ref{eq:G})}
		\State $O(n) = (L^*_{tr} - L_{tr}(n)) - (\bar{L}^*_{\diamond} - \bar{L}_{\diamond}(n))$ \Comment{Eq. (\ref{eq:O})}
		\State $w(n) = \frac{1}{Z}\frac{G(n)}{O^2(n)}$  \Comment{Eq. (\ref{eq:w})}
		\State{\bf if~}{$\bar{L}_{tr} < L^*_{tr}$}{\bf ~then~}$L^*_{tr} = \bar{L}_{tr}$ \Comment{Update best training loss}
		\State{\bf if~}{$\bar{L}_{\diamond} < L^*_{\diamond}$}{\bf ~then~}$L^*_{\diamond} = \bar{L}_{\diamond}$ \Comment{Update best true loss}
		\EndProcedure
	\end{algorithmic}
\vspace{-0.15cm}
\end{algorithm}

Equations (\ref{eq:G}) and (\ref{eq:O}) suggest that the accuracy of the approximations for $G^{(k)}$ and $O^{(k)}$ depends on the references $L^{*(k)}_{tr}$ and $L^{*(k)}_\diamond$. In \cite{Wang2020,Phan2020}, the losses $L^{(k)}_{tr}(0)$ and $L^{(k)}_\diamond(0)$ at time $n=0$ (i.e. right after the network initialized with random weights) were used for this purpose. However, we empirically found that these fixed references resulted in unsatisfactory performance. We conjecture that it is most likely due to the bias to a specific random initialization of the network (i.e. different random initializations will lead to various approximation accuracy). To overcome this, we propose to use the best losses up the current time $n$ for references. Furthermore, these references are also adapted during the training course. Furthermore, in audio classification tasks, even though the overall trend of the loss curves are smooth, they are noisy in short term, we therefore smooth the losses with a history window of size $W$ before updating the loss references to avoid being stuck in local minima. The procedure for computing the weight $w^{(k)}$ is devised in Algorithm~1.

\vspace{-0.15cm}
\subsection{Self-ensemble}
\vspace{-0.15cm}
Since the multi-view network has multiple outputs (i.e. the single-view classification outputs and the multi-view classification output) which can be aggregated to produce a self-ensemble of decisions:
\begin{align}
P(y = c)=\frac{1}{5}\sum_{k} \left(w^{(k)}_{\diamond}P^{(k)}(y = c)\right),
\end{align}
where $P^{(k)}(y = c)$ denotes the probability that the classification branch $k$ predicts the category $c \in \{1, \ldots, C\}$ out of $C$ categories. We use $w^{(k)}_\diamond$ to denote the  weight of the classification branch $k$ found with the final model. The final output label is then determined as:
\begin{align}
\hat{y} = \argmax_{c} P(y = c).
\end{align}
\vspace{-0.55cm}
\section{Experiments}
\label{sec:experiments}
\vspace{-0.15cm}
\subsection{Experimental setup}
\label{ssec:experimental_setup}

\vspace{-0.1cm}
\subsubsection{Datasets}
\vspace{-0.1cm}
We employed three databases to conduct experiments on three audio and music classification tasks: environmental sound classification, audio scene classification, and music genre classification.

\begin{figure*} [!t]
	\centering
	\includegraphics[width=0.8\linewidth]{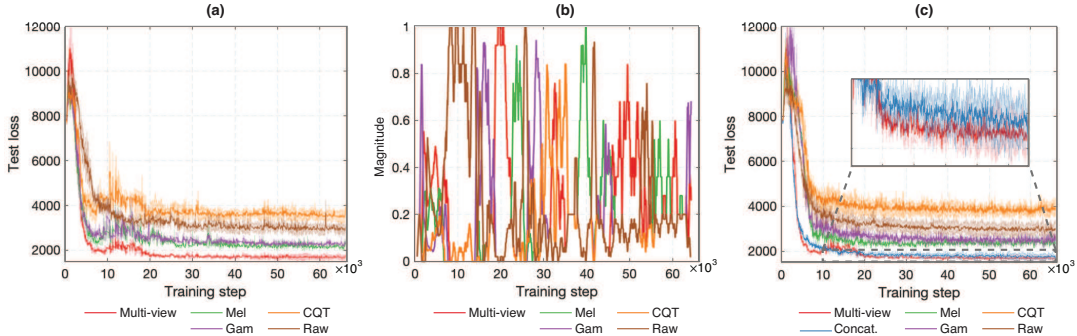}
	\vspace{-0.4cm}
	\caption{\small ESC-50: (a) The test loss curves (averaged over 5 cross-validation folds) of the classification branches of the multi-view network; (b) The weights assigned to the classification branches of the multi-view network during training (only the first cross-validation fold is shown); (c) The test loss curves (averaged over 5 cross-validation folds) of the multi-view network and the baselines during training.}
	\label{fig:loss_weight}
		\vspace{-0.3cm}
\end{figure*}

{\bf ESC-50 \cite{Piczak2015}:} This dataset consists of 2,000 monaural samples in total which are evenly distributed among 50 environmental sound categories. Each sample has a length of roughly 5 seconds sampled at 44.1 kHz. The dataset was divided into 5 folds and we adhered to \cite{Piczak2015} to conduct 5-fold cross validation.

{\bf DCASE2016 Task 1 \cite{Mesaros2016}:} This dataset was used in the audio scene classification (Task 1) of the DCASE 2016 challenge \cite{Mesaros2016}. It consists of 1,560 binaural samples evenly distributed among 16 audio scene categories. The data was recorded with a sampling frequency of 44.1 kHz and each sample has a length of 30 seconds. We used the development set for training and the evaluation set for testing in the experiments. Note that, for simplicity, binaural audio was reduced to monaural before experimentation.

{\bf GTZAN \cite{Tzanetakis2002}:} This dataset has been widely used for evaluation of music genre classification. It consists of 10 genres with 100 audio files each, all having a length of 30 seconds and sampling frequency of 22,050 Hz. We conducted 10-fold cross validation following \cite{Tzanetakis2002}.

\vspace{-0.1cm}
\subsubsection{Feature extraction}
\vspace{-0.1cm}
To extract the 2D low-level features, a raw audio signal was transformed into a log Mel-scale spectrogram using $F = 64$ Mel-scale filters in the frequency range up to Nyquist rate. Similarly, log Gammatone spectrogram was extracted using $F = 64$ Gammatone filters. A window size of 40ms and 50\% overlap were commonly used. Log CQT spectrogram \cite{Schoerkhuber2010} was extracted using Librosa \cite{McFee2015} with $F=64$ frequency bins, 12 bins per octave, and a hop length of 512 (for 22,050 Hz sampling rate) or 1024 (for 44.1 kHz sampling rate). 

Note that with this setting, the time dimension of the CQT spectrogram is smaller than that of the Mel-scale and Gammatone ones. A 30-second snippet at 44.1 kHz sampling rate results in a Mel-scale and a Gammatone spectrogram of size $1499 \times64$ while the resulted CQT spectrogram is of size $1292\times64$.

\vspace{-0.1cm}
\subsubsection{Parameters}
\vspace{-0.1cm}
The network was trained using Adam optimizer \cite{Kingma2015} for $E\!=\!3000$ epochs (ESC-50) and $E\!=\!1500$ epochs (DCASE 2016 and GTZAN) with a minibatch size of 64. 
We used the Mel-scaled and Gammatone inputs of length $T\!=\!75$ frames, the CQT input of length of $T\!=\!65$ frames, and the raw waveform input of length $66,650$ samples (with 44.1 kHz sampling rate) or 33,330 samples (with 22,050 Hz sampling rate). It should be noted that when the raw input has length of 33,330, the \emph{pool02} layer in the 1D CRNN (cf. Table \ref{tab:1dcrnn}) had its kernel size and stride reduced by half.

The learning rate was initially set to $2\times10^{-4}$ and was exponentially reduced with a rate of 0.8 after $0.1E$, $0.2E$, and $0.3E$ epochs. In addition, the first 10 epochs were used as a warm-up period in which the network was trained with a small learning rate of $2\times10^{-5}$. For model selection and for approximating the true loss in (\ref{eq:G}) and (\ref{eq:O}), a validation set was randomly drawn and left out. More specifically, samples from two audio sources per category in case of ECS-50, samples from 10\% of audio sources per category in case of DCASE2016, and 10\% of samples in case of GTZAN were used for this purpose. During training, the network that resulted in best validation accuracy was retained. Note that, in order to compute the training loss in (\ref{eq:O}), evaluating the network on the entire training set would be computationally expensive. Instead, we sampled and fixed a small subset of training examples (roughly the same size as the validation set) for approximation.

During testing, for an audio file of length $S$ seconds, $S$ data samples were evenly sampled and presented to the trained network for classification. The global classification decision was obtained by aggregating the segment-wise decisions via averaging.

\vspace{-0.1cm}
\subsubsection{Baselines}
\vspace{-0.1cm}
To assess the efficacy of the proposed multi-view method, we constructed four single-view baselines and two multi-view baselines for comparison. The four single-view baselines were the CRNN subnetworks in Fig. \ref{fig:multiview_network} that were trained independently on the individual low-level inputs. The first multi-view baseline had a similar architecture to the proposed multi-view network (cf. Fig. \ref{fig:multiview_network}) but relied on simple concatenation fusion. The second multi-view baseline was a late-fusion system that combined the independent single-view baselines by taking average of their classification probabilities. 

\vspace{-0.1cm}
\subsection{Experimental results}
\vspace{-0.1cm}
The classification accuracies obtained by the proposed multi-view method and the baselines over the experimental databases are shown in Table~\ref{tab:results}. On the one hand, among the single-view baselines, the ones using the Mel-scale and Gammatone spectrogram inputs results in better performance than those using the CQT spectrogram and raw inputs. This result is consistent with the finding in \cite{Huzaifah2017} and also reflects the fact that the former two are more popular than the latter two in various audio/music recognition tasks. Although combining multiple views via the simple concatenation and late fusion leads to performance gains in all the experimental databases, it is disputable whether late fusion works better than concatenation since the former outperforms the latter on ESC-50 and GTZAN whereas the opposite result was seen on DCASE2016.

On the other hand, the proposed multi-view network consistently outperforms not only the single-view baselines but also the multi-view baselines over all three tasks. More specifically, our network achieves an accuracy gain of $0.85\%$, $1.28\%$, and $1.1\%$ absolute on ESC-50, DCASE2016, and GTZAN over the best baseline (i.e. late fusion, concatenation, and late fusion), respectively. The superiority of the proposed method is also reflected by its lower test loss as shown in Fig.~\ref{fig:loss_weight}\,(c). The gain via self-ensembling is even better, achieving $1.55\%$, $1.53\%$, and $1.4\%$ absolute, respectively. 

These results suggest that the proposed multi-view learning method is more efficient than the popular concatenation and late fusion methods. It can be explained that the proposed method offers a mechanism to harmonize learning rhythms of the individual views and cohere them to consolidate the joint representation. This mechanism is partly illustrated in Fig.~\ref{fig:loss_weight}\,(a) and (b), particularly from the training step 0 to 10,500. In this period, the 2D branches were converging faster than the 1D branch and were given higher weights until they started degenerating around the training step 7,500. The weights for the 2D branches were then reduced to slow down their learning while the weight for the 1D branch, which was still converging well at the time, was steadily increased to accelerate its learning. It is most likely that such a mechanism is lacking in the simple concatenation whereas late fusion of the single-view models is suboptimal as separate training ruled out cross-view interaction. 

%\subsection{Discussion}

%\textcolor{red}{We did not tailored the network to different tasks or explored channel combinations. But it could be done.}

It should be emphasized that our primary goal in this work is to study and compare the proposed multi-view learning method to the common multi-view fusion methods with respect to a fixed network architecture rather than a comprehensive performance comparison with existing works. We, therefore, neither tailored the network architecture for the individual tasks \cite{Wang2020a, Liu2019, Guzhov2020} nor explored multi-channel combinations \cite{Han2016, Liu2019, Pham2020}. Readers should be informed that, for the databases we adopted in this study, better performance was reported in other works, such as \cite{Wang2020a, Guzhov2020} for ESC-50, \cite{Pham2020, Yin2018} for DCASE2016, and \cite{Liu2019} for GTZAN. Incorporating the propose multi-view method to existing state-of-the-art networks is worth further investigation.

\setlength\tabcolsep{1.25pt}
\begin{table}[t!]
	\caption{\small Results obtained by the studied speech enhancement systems on the objective evaluation metrics.}
		\vspace{-0.2cm}
	\begin{center}
		\footnotesize
		\begin{tabular}{|>{\arraybackslash}m{0.8in}|>{\centering\arraybackslash}m{0.5in}|>{\centering\arraybackslash}m{0.7in}|>{\centering\arraybackslash}m{0.5in}|}
			\cline{2-4}
			\multicolumn{1}{c|}{}& ESC-50 & DCASE2016 & GTZAN  \parbox{0pt}{\rule{0.pt}{0ex+\baselineskip}}\\ [0ex] 	% inserts table heading
			\hline
			\bf Self-ensemble & $\mathbf{87.35}$  & $\mathbf{85.38}$  & $\mathbf{91.10}$   \parbox{0pt}{\rule{0.pt}{0ex+\baselineskip}}\\ [0ex] 	% inserts table heading
			\bf Multi-view & $\mathbf{86.65}$  & $\mathbf{85.13}$  &  $\mathbf{90.80}$  \parbox{0pt}{\rule{0.pt}{0ex+\baselineskip}}\\ [0ex] 	% inserts table heading
			%\emph{Mel} & $\it 81.30$  & $\it 81.54$ &  $\it 87.70$ \parbox{0pt}{\rule{0.pt}{0ex+\baselineskip}}\\ [0ex] 	% inserts table heading
			%\emph{Gammatone} & $\it 83.59$  & $\it 76.92$  &  $\it 86.80$  \parbox{0pt}{\rule{0.pt}{0ex+\baselineskip}}\\ [0ex] 	% inserts table heading
			%\it CQT & $\it 59.75$  & $\it 74.36$ &  $\it 83.70$  \parbox{0pt}{\rule{0.pt}{0ex+\baselineskip}}\\ [0ex] 	% inserts table heading
			%\it Raw & $\it 73.70$  & $\it 45.13$ &  $\it 77.80$  \parbox{0pt}{\rule{0.pt}{0ex+\baselineskip}}\\ [0ex] 	% inserts table heading
			\hline
			Late fusion & $85.80$ & $82.05$  &  $89.70$ \parbox{0pt}{\rule{0.pt}{0ex+\baselineskip}}\\ [0ex] 	% inserts table heading
			Concat. fusion & $84.44$  & $83.85$  &  $89.00$ \parbox{0pt}{\rule{0.pt}{0ex+\baselineskip}}\\ [0ex] 	% inserts table heading
			Mel & $80.15$  & $77.18$ &  $87.30$ \parbox{0pt}{\rule{0.pt}{0ex+\baselineskip}}\\ [0ex] 	% inserts table heading
			Gammatone & $76.90$  & $77.95$  &  $86.20$ \parbox{0pt}{\rule{0.pt}{0ex+\baselineskip}}\\ [0ex] 	% inserts table heading
			CQT & $58.30$  & $75.38$ &  $83.10$ \parbox{0pt}{\rule{0.pt}{0ex+\baselineskip}}\\ [0ex] 	% inserts table heading
			Raw & $75.60$  & $57.44$  &  $81.50$ \parbox{0pt}{\rule{0.pt}{0ex+\baselineskip}}\\ [0ex] 	% inserts table heading
			\hline
		\end{tabular}
	\end{center}
	\label{tab:results}
	\vspace{-0.6cm}
\end{table}

\vspace{-0.15cm}
\section{Conclusions}
\vspace{-0.15cm}
We presented in this work a novel multi-view learning approach for audio and music classification. The proposed multi-view network was designed to have multiple CRNN subnetworks, each handling one input view. The multi-view embedding was then produced by concatenating the embeddings learned by the single-view subnetworks. In addition to the classification branch on the multi-view embedding, the network also accommodated classification branches on the single-view subnetworks that offered a means to assess their learning behavior. Each classification branch was assigned to a weight that was adapted during training to reflect whether it is generalizing well or overfitting the data. The gradients from different classification branches were blended according to their weights for network training. In this way, different views were supposed to contribute proportionally to the multi-view embedding depending on their learning behavior. The efficacy of the proposed method was demonstrated in three different audio and music classification tasks on which the proposed method outperformed all the single-view and multi-view baselines.

	\small
	%\pagebreak
	\bibliographystyle{IEEEbib}
	\bibliography{reference}
\end{document}